\documentclass[12pt]{article}
\usepackage{a4wide,graphicx,natbib}
\begin{document}
\thispagestyle{empty}

\begin{center}

\begin{LARGE}
\centerline{\bf Surprises in approximating Levenshtein distances}
\end{LARGE}
\bigskip\bigskip\bigskip

\begin{large}
Michael~Baake\,$^{\rm a}$, 
Uwe~Grimm\,$^{\rm b}$~and Robert~Giegerich\,$^{\rm c}$
\end{large}
\bigskip

\begin{footnotesize}
$^{\rm a}$Fakult\"{a}t f\"{u}r Mathematik, Universit\"{a}t Bielefeld,
Postfach 100131, 33501 Bielefeld, Germany\\
$^{\rm b}$Department of Mathematics, The Open University, 
Milton Keynes MK7 6AA, UK\\
$^{\rm c}$Technische Fakult\"{a}t, Universit\"{a}t Bielefeld,
Postfach 100131, 33501 Bielefeld, Germany\\
\end{footnotesize}
\end{center}
\bigskip

\begin{abstract}
  The Levenshtein distance is an important tool for the comparison of
  symbolic sequences, with many appearances in genome research,
  linguistics and other areas. For efficient applications, an
  approximation by a distance of smaller computational complexity is
  highly desirable.  However, our comparison of the Levenshtein with a
  generic dictionary-based distance indicates their statistical
  independence.  This suggests that a simplification along this
  line might not be possible without restricting the class of
  sequences. Several other probabilistic properties are briefly
  discussed, emphasizing various questions that deserve further
  investigation.
\end{abstract}
\bigskip

\centerline{\footnotesize\textbf{Keywords:} 
global alignment; distance concepts; statistical independence; 
computational complexity}


\bigskip
\bigskip
\section{Introduction}

The Levenshtein (or edit) metric \citep{Leven65} is a standard tool to
estimate the distance between two sequences. It is widely used in
linguistics and bioinformatics, and for the recognition of text blocks
with isolated mistakes. As is well known, its computational
complexity, when applied to two sequences of (approximately) the same
length $n$, is $\mathcal{O}(n^{2})$. Since this is a hurdle in many
practical applications, it is desirable to replace, or to approximate,
the Leven\-shtein (L) distance by some quantity of smaller (preferably
linear) computational complexity.  Two fast approximation algorithms
for edit distances were suggested by \cite{U}, one based on maximal
exact matches, the other on suitably restricted subword comparisons
between the two sequences; compare also \cite{Lipp02}.  This would
indeed give $\mathcal{O}(n)$, due to their computability from the
suffix tree; see \cite{Gus}. However, they only provide lower bounds,
and hence no complete solution of the problem.

It seems possible to estimate probabilistically, with sublinear
complexity, whether the L-distance of two sequences is `small' or
`large'; see \cite{Batu03}.  Whether an improvement of this rather
coarse result or even a replacement of the L-distance is possible,
with at most linear complexity and a non-probabilistic outcome, seems
open. Below, we compare the L-distance with a representative
dictionary-based distance. Our findings support the conclusion that
such a simplification might be difficult or even impossible. On the
way, we highlight some interesting properties that have been neglected
so far, but seem relevant for a better understanding of such distance
concepts.

\section{Comparison of two distances}

To keep discussion and results transparent, we concentrate on two
specific distances, and on binary sequences.  We have also tried a
number of obvious alternatives, but they did not show any
significantly different behaviour. In this sense, the structure of our
example is more likely typical than exceptional.

The L-distance $d_{\mathrm{L}}(u,v)$ of two sequences $u$ and $v$ (not
necessarily of equal length) is the minimum number of edit operations
(insertions, deletions, or substitutions) needed to transform $u$ into
$v$ or vice versa \cite[Ch.~11.2]{Gus}. Though $d_{\mathrm{L}}(u,v)$
is closely related to the longest common subsequence (LCS) (loc.\
cit., Ch.~11.6.2) of $u$ and $v$ (and hence to distances based upon
it), one important difference lies in the possibility of
substitutions. So, using the LCS in this context requires some
care. For sequences of lengths $m$ and $n$, the computational
complexity of calculating $d_{\mathrm{L}}$ (or the LCS) is
$\mathcal{O}(mn)$, e.g., when based on the Needleman-Wunsch algorithm;
see \cite[Ch.~6.4.2]{EG}.

A generic choice for a dictionary-based metric is
\[
   d_{\mathrm{D}}(u,v) = \mathrm{card}\big(\mathcal{A}(u) 
      \triangle \,\mathcal{A}(v)\big)\, ,
\]
where $\mathcal{A}(u)$ is the full dictionary of $u$, i.e., the set of
all non-empty subwords of $u$, and $A\triangle B=(A\cup
B)\setminus(A\cap B)$ is the symmetric difference of $A$ and $B$.
This choice actually disregards the goal of computational
simplification, but focuses on the full dictionary information
instead, and thus, in some sense, represents the optimal information
on the sequences to be compared.  It is well known that, using the
suffix tree structure, the calculation of closely related
dictionary-based distances is possible with linear complexity, e.g.,
by means of Ukkonen's algorithm; compare \cite[Ch.~6]{Gus}. On the
other hand, further restrictions are likely to reduce the usefulness
in relation to the L-distance.

Both $d_{\mathrm{L}}$ and $d_{\mathrm{D}}$ define a {\em metric},
i.e., for arbitrary sequences $u,v$ and $w$, the distance
$d\in\{d_{\mathrm{L}},d_{\mathrm{D}}\}$ satisfies the axioms of a
metric \cite[Ch.~2.11]{Schechter}:
\begin{itemize}
\item[(i)] $0\le d(u,v)<\infty$ (positivity);
\item[(ii)] $d(u,v)=0$ if and only if $u=v$ (non-degeneracy); 
\item[(iii)] $d(u,v)=d(v,u)$ (symmetry); 
\item[(iv)] $d(u,v)\le d(u,w)+d(w,v)$ (triangle inequality).
\end{itemize}

Less clear is the relation between $d_{\mathrm{L}}$ and
$d_{\mathrm{D}}$. Since one can easily construct pairs of sequences
that are close in one, but not in the other distance, they are
certainly not equivalent in the strong sense as also used for norms,
compare \cite[Ch.~I.2]{Werner}. They are equivalent in the weaker
sense of generating the same topology \cite[Ch.~22.5]{Schechter},
which is the discrete topology here. However, this is of little use
for the question addressed above. The situation does not improve if
one replaces $d(u,v)$ by the quotient $d(u,v)/(1+d(u,v))$, which is
another metric, with range in $[0,1]$. As we shall see below, the
situation is actually much worse.

\section{Concrete results}

\begin{figure}
\centerline{\includegraphics[width=0.8\textwidth]{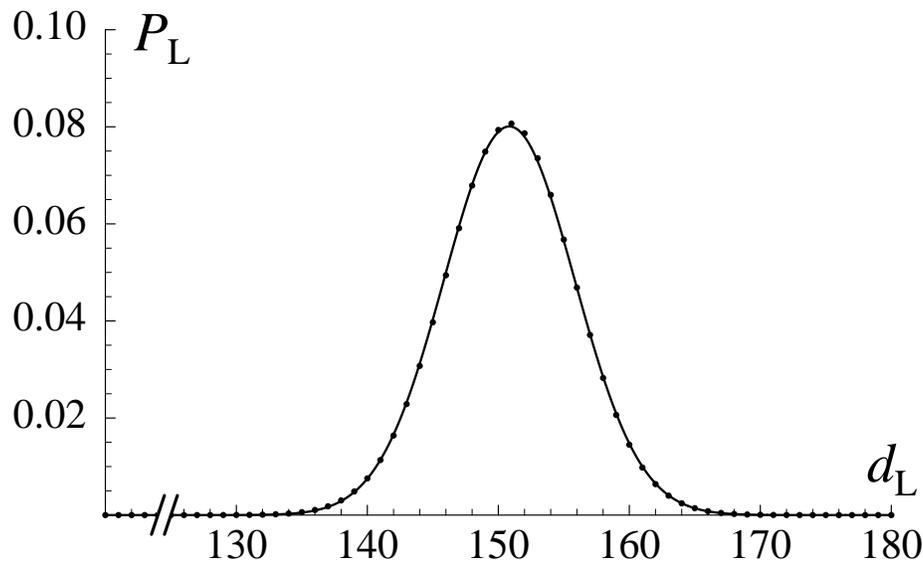}}
\caption{Simulated probability distribution $P_{\mathrm{L}}$ of the 
L-distance $d_{\mathrm{L}}$ between two random
sequences of length $500$ (dots) and Gaussian approximation
(line), with mean $150.84$ and variance $24.82$.}\label{gauss}
\end{figure}

To get a first impression of the L-distance, we computed the discrete
probability distribution of the values $d_{\mathrm{L}}(u,v)$ for
sequences $u\ne v$ of the same length, under uniform distribution on
sequence space. This has long been known to be a reasonable first
approach for the comparison of sequences from data bases \citep{RDD}.
Up to length 20, this was done using all possible pairs; for longer
sequences, the distribution was estimated from a sufficiently large
random selection of pairs. For length $n=500$, the result obtained
from $4\!\times\! 10^8$ pairs is shown in Figure~\ref{gauss}.  For
large $n$, the distributions seem to be well described by Gaussian (or
normal) distributions. This qualitative behaviour does not change much
and seems to improve with sequence length. One could add weight to
this finding by performing a statistical test on Gaussianity, which
would score well. However, we think that one should not over-interpret
this observation, in particular in view of a recent numerical
investigation by \cite{PTCT} which indicates that a gamma distribution
might give an even better description.

Note that, if extrema over {\em local}\/ alignments are taken, one
obtains an extremal value distribution \cite[Ch.~2.3.2]{PW}. However,
this implies nothing for the {\em global}\/ alignment considered here.
The possible (or approximate) Gaussian nature of this case has been
observed before by Dayhoff, see \cite[Ch.~3]{Mount} and references
given there; a more detailed investigation of tail probabilities can
be found in \cite{Waterman}.  Still, it seems to be hardly noted,
although it is a relevant phenomenon that deserved further attention,
with exact results presently not in sight.

\begin{figure}
\centerline{\includegraphics[width=0.8\textwidth]{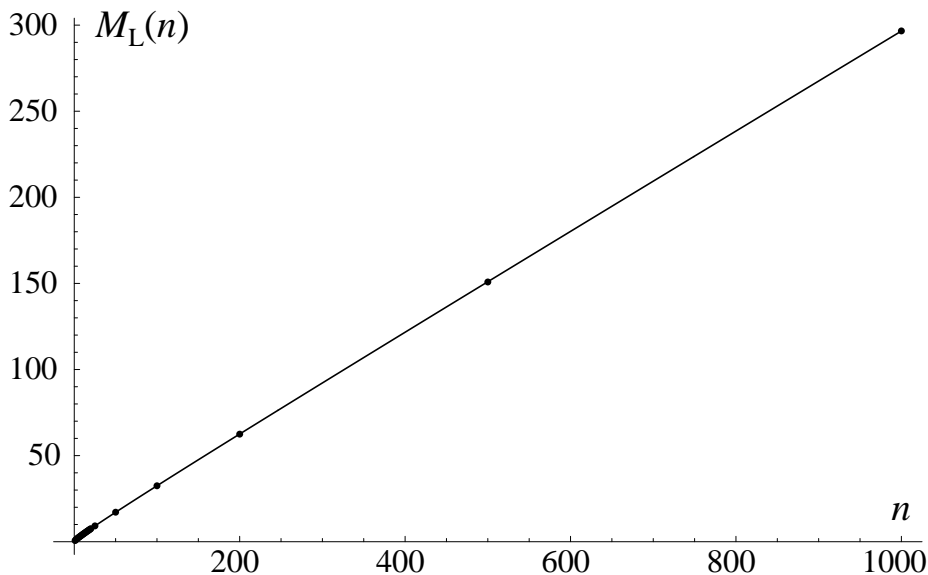}}\smallskip
\caption{Mean $M_{\mathrm{L}}(n)$ of the probability distribution 
$P_{\mathrm{L}}$ as a function 
of sequence length $n$, calculated exactly for $n\le 20$ and by simulation 
otherwise. The solid line shows the least squares fit 
$M_{\mathrm{L}}(n)=0.413\sqrt{n}+0.283\, n$.}
\label{meanplot}
\end{figure}

\begin{figure}
\centerline{\includegraphics[width=0.8\textwidth]{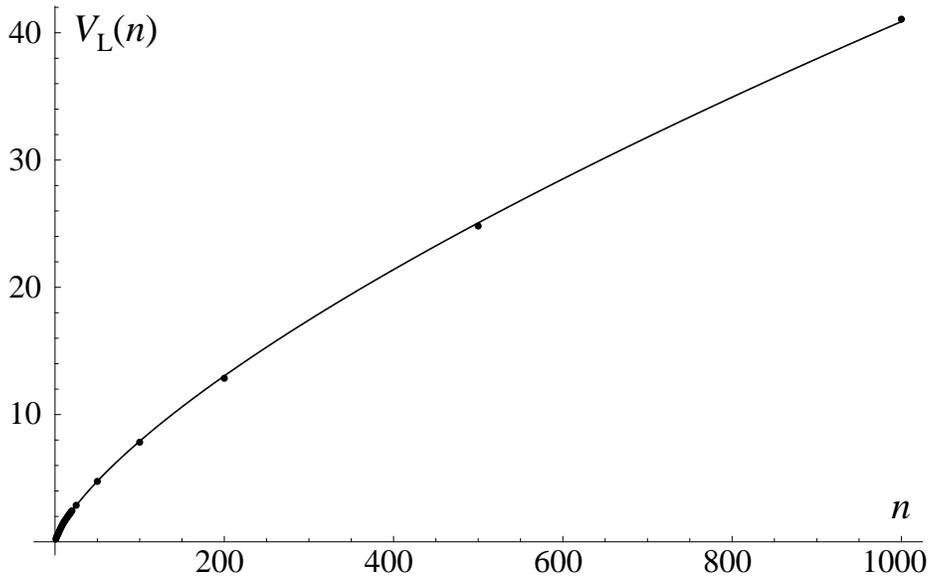}}\smallskip
\caption{Variance $V_{\mathrm{L}}(n)$ of the probability distribution 
$P_{\mathrm{L}}$ as a function 
of sequence length $n$, calculated exactly for $n\le 20$ and by simulation 
otherwise. The solid line shows the least squares fit 
$V_{\mathrm{L}}(n)=-0.283\sqrt{n}+0.498\, n^{2/3}$.}
\label{varplot}
\end{figure}

For this reason, we could only investigate our findings numerically.
Beyond checking the Gaussian behaviour qualitatively, means and
variances were calculated for different $n$, both by exact
enumeration (for $n\le 20$) and by simulation (for larger $n$, up to
$n=1000$). It is an interesting question whether the mean and the
variance, as functions of sequence length, show power-law behaviour,
at least asymptotically. Our data, see Figures \ref{meanplot} and
\ref{varplot}, are compatible with an asymptotically linear growth of
the mean and an asymptotic $n^{2/3}$ power law for the variance, both
with a square-root correction term (for which we do not have any
particular justification).  Such predictions and conjectures are
presently discussed by various people \citep{Matz}. In particular, the
$n^{2/3}$ power law for the variance would be in line with analogous
observations for the LCS, compare \cite{HL}. Since there has recently
been some doubt in the correctness of this finding \citep{Matz}, it
requires further corroboration and investigation.

A similar finding (though with larger fluctuations) applies to the
distribution of the values $d_{\mathrm{D}}(u,v)$ for random pairs
$u\ne v$. However, there is no compelling reason to investigate this
specific distance in detail, as it was mainly selected for
illustrative purposes and does not seem to be closely related to one
of the standard problems of probability theory.

\begin{figure}[ht]
\centerline{\includegraphics[width=0.8\textwidth]{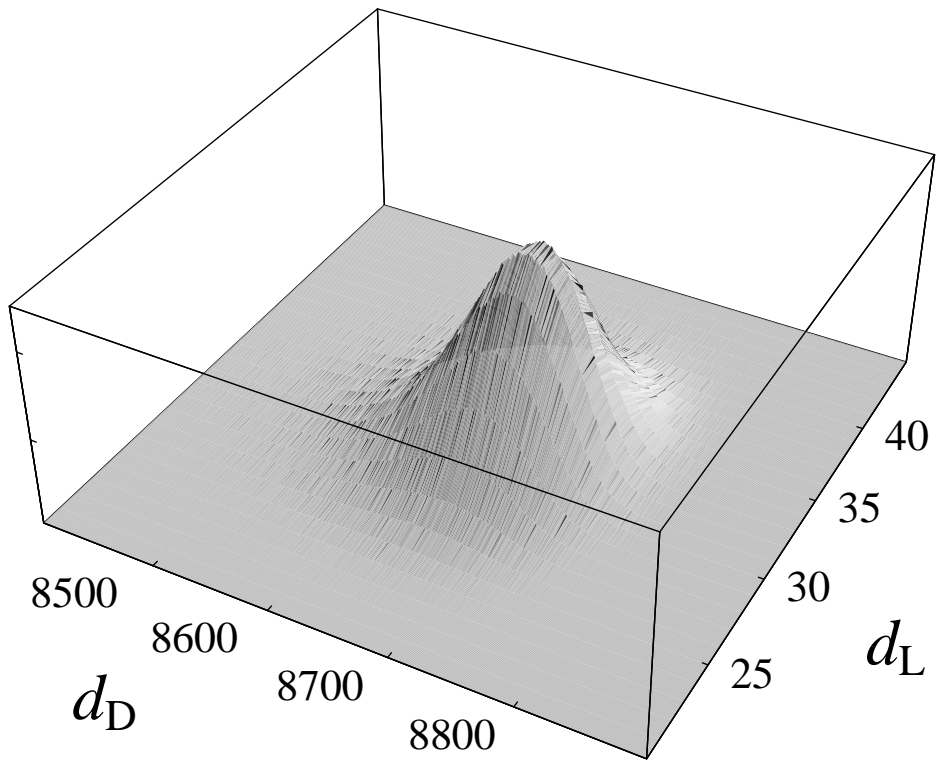}}\smallskip
\caption{Numerical approximation of the joint probability distribution
$P$ for $d_{\mathrm{D}}$ and $d_{\mathrm{L}}$, obtained from a
simulation with $10^7$ random pairs of sequences of length $100$. Both 
marginal distributions are approximately Gaussian, with slightly
larger fluctuations for the distribution of 
$d_{\mathrm{D}}$-values.}
\label{corr}
\end{figure}

More interesting, and also more relevant, is the question for the {\em
joint}\/ distribution of $d_{\mathrm{D}}(u,v)$ and
$d_{\mathrm{L}}(u,v)$. A necessary requirement for a useful relation
between the two distances would be a strong correlation.  However, as
Figure~\ref{corr} shows for sequences of length $100$, there is little
correlation at all -- the joint distribution is rather well described
by the product of the two Gaussians needed for the marginal
distributions. This observation could be quantified with some effort,
but we refrain from doing so because it would not contribute to the
interpretation at this stage. 

Our finding means that, at least on the level of the full sequence
space or for the alignment of two random sequences (as analyzed in our
simulations), the distances $d_{\mathrm{D}}(u,v)$ and
$d_{\mathrm{L}}(u,v)$ are closer to being statistically independent of
each other than to being useful approximations of one another.

\section{Concluding remarks}

Our findings are to be interpreted with care. They do not rule out a
simplified approach to L-type distances, at least when restricted to
(possibly relevant) subsets of sequences. However, they seem to
indicate that subword comparison leads to statistically independent
information, at least when viewed on the full sequence space.
Clearly, different distance concepts can and should be tried.
Moreover, a rigorous stochastic analysis of the various limit
distributions is necessary to clarify the picture obtained from the
simulations.

As long as analytic results (e.g., via limit theorems) are
unavailable, it would also help to perform a more detailed statistical
analysis of the various distributions, including clear-cut statistical
tests. In particular, it would be extremely relevant to also consider
suitable subspaces of the full sequence space, such as those
extractable from existing data bases. Though this is clearly far
beyond the scope of this short note, we believe that it would be a
rewarding task for future investigations.

\section*{Acknowledgements}

It is our pleasure to thank E.~Baake and D.~Lenz for helpful
discussions, and F.\ Merkl and M.\ Vingron for useful hints on the
literature. Financial support from British Council (ARC 1213) and
DAAD is gratefully acknowledged.

\end{document}